# LOW LEVEL RF CONTROL FOR THE PIP-II ACCELERATOR

J. P. Edelen[#], B. E. Chase, E. Cullerton, J. Einstein-Curtis, J. Holzbauer, D. Klepec, Y. Pischalnikov, W. Schappert, P. Varghese, Fermilab, Batavia, IL

G. Joshi, S. Khole, D. Sharma, Department of Atomic Energy, India

*Abstract*

The PIP-II accelerator is a proposed upgrade to the Fermilab accelerator complex that will replace the existing, 400 MeV room temperature LINAC with an 800 MeV superconducting LINAC. Part of this upgrade includes a new injection scheme into the booster that levies tight requirements on the LLRF control system for the cavities. In this paper we discuss the challenges of the PIP-II accelerator and the present status of the LLRF system for this project.

## INTRODUCTION

In order to facilitate the next generation of neutrino experiments, Fermilab has proposed an upgrade to the accelerator complex to enable operation with a beam power of 1.2 MW. This upgrade includes the construction of an 800 MeV superconducting H- linear accelerator [1]. Figure 1 shows the proposed PIP-II LINAC on the infield of the old Tevatron/Main Ring and how the transfer line will be constructed to bring the beam from the LINAC into the booster.

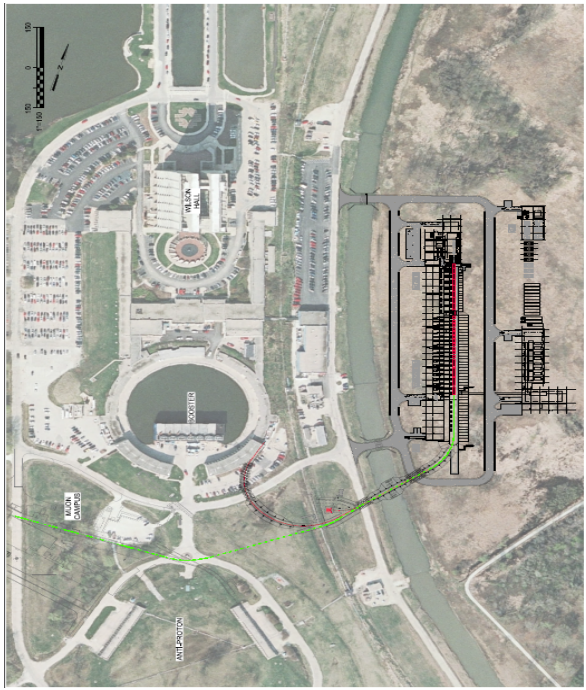

Figure 1: High level view of how the PIP-II LINAC fits into the Fermilab accelerator complex

As part of this upgrade, injection into the booster synchrotron will be handled by longitudinal phase space painting in order to improve the longitudinal stability of the beam [ibid]. In order for this scheme to perform as desired, the beam generated in the LINAC must have very tight tolerances on the energy spread that is also stable with time [ibid]. This levies a requirement on the LLRF system to regulate the amplitude and phase of each cavity to better than $0.01°$ in phase and $0.01\%$ in amplitude. This level of regulation is challenging, but achievable using state-of-the art practices for LLRF system design. Additionally the LLRF system is responsible for the generation of a multi-frequency Master Oscillator and Phase Reference lines, Beam Chopper Waveform Generator, RF locking source for the booster ring during beam fill, RF synchronous timing source, and superconducting RF resonance control (microphonics and LFD). In this paper we begin with an overview of the PIP-II LINAC, discuss simulation efforts to understand the sensitivity of the LINAC to static errors, provide an update for the LLRF system design, and then present status of the hardware development.

## OVERVIEW OF PIP-II

The PIP-II LINAC is a 20 Hz, 800 MeV, superconducting H- LINAC that will replace the existing 400 MeV copper LINAC. The primary goal of PIP-II is to increase the available proton power for neutrino experiments to 1.2 MW. As part of this upgrade, we are constructing a full test of the warm front end through the second superconducting cryomodule. The primary goal of this test is to test the ability to chop the beam produced in the RFQ to the correct bunch pattern for synchronous transfer from the LINAC to the booster, via phase space painting. The secondary goal is to test the transition from the warm front end to the superconducting LINAC and prove out other high-risk technology.

The RF systems for the LINAC are broken down into three frequency sections, 162.5 MHz, 325 MHz, and 650 MHz. The 162.5 MHz section is comprised of three cavity types, the RFQ, four bunching cavities, and eight superconducting half wave resonators. The RFQ is powered by two 75 kW solid state amplifiers, the bunching cavities are each powered individually by a single 3 kW solid state amplifier, and the half wave resonators are powered individually by 7 kW solid state amplifiers. All of the power systems in the 162.5 MHz section are operated in CW mode. The 325 MHz section consists of two types of superconducting single spoke resonators. The first type is powered by pulsed amplifiers that operate at 7 kW while the second type is powered by 20 kW amplifiers. The 650 MHz section consists of two types of superconducting elliptical cavities powered by 40 and 70 kW amplifiers for the low beta and high beta sections respectively. Figure 2 shows a cartoon of the different cavities in the PIP-II LINAC and their respective frequencies.

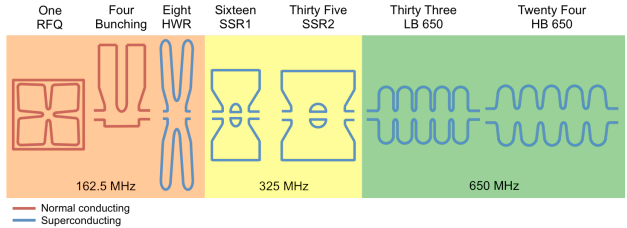

Figure 2: Cartoon of the PIP-II LINAC

## LINAC ENERGY STABILITY SIMUALTIONS

In order to better understand our specifications, we studied the sensitivity of the beam energy and phase at the end of the LINAC to perturbations in the cavity phase, Figure 3. This shows that in general the individual cavities do not have a large impact on the output energy. Aggregate errors have also been studied and these are the source of the amplitude and phase performance specifications. The tools developed to perform these sensitivity studies have been improved and now incorporate RF transient effects and dynamic errors during the RF pulse. A paper on this tool will be published in the near future.

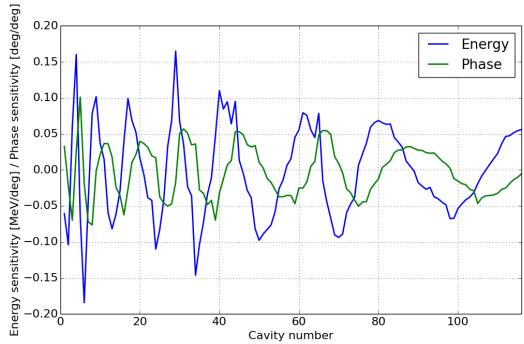

Figure 3: Sensitivity of the beam energy and phase to perturbations in individual cavity phases

We also studied the effect of phase errors that occur between frequency sections due to errors in the phase reference system. For this three cases were examined: A one degree perturbation in the 325 MHz system, a one degree perpetuation in the 650 MHz system, and a compounding error where there is one degree error produced in the 325 MHz system that leads to a two degree error in the 650 MHz system. Figure 4 shows the beam energy error as a function of position along the LINAC for these different scenarios.

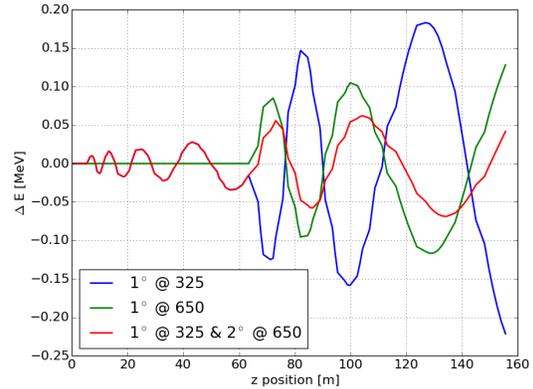

Figure 4: Energy sensitivity to phase errors introduced by frequency transitions

Here we see that by far the worst situation is the most unlikely, where there is a phase error at the 325 MHz interface but no error at the 650 MHz interface. The case with a phase error only at the 650 MHz interface is almost as bad, while the compounding errors are the least bad. In all cases these errors produce energy errors that are unacceptable therefore we need be sure that the different frequency sections are properly aligned to the reference.

## RESONANCE CONTROL

The resonance control specifications continue to be the largest challenge for the successful operation of PIP-II. For each cavity type the total detuning must during the flat-top must be less than 25 Hz. This is a very tight requirement considering there are both microphonic disturbances and Lorentz Force Detuning. The Lorentz Force Detuning is particularly challenging for these cavities because they are very narrow band and operate at 20 Hz. Table 1 shows a summary of the LFD and df/dp measurements completed to date.

Table 1: Summary of detuning specifications on the cavities

|  | HWR | SSR1 | SSR2 | LB650 | HB650 |
|---|---|---|---|---|---|
| Sensitivity to He pressure (FRS), $df/dP$, Hz/Torr | <25 | <25 | <25 | <25 | <25 |
| … (measurements), $df/dP$, Hz/Torr | 13 | 4.0 | - | - | - |
| Estimated LFD sensitivity, $df/dE^2$, Hz/(MV/m)$^2$ | - | -5.0 | - | -0.8 | -0.5 |
| … (measurements), $df/dE^2$, Hz/(MV/m)$^2$ | -1.5[*] | -4.4 | - | - | - |
| Estimated LFD at nominal voltage (FRS), Hz | - | -500 | - | -192 | -136 |
| … (measurements) at nominal voltage, Hz | -122.4 | -440 | - | - | - |

[*] Two cavities were measured in a test stand. The results are: -1.82 and -1.3 Hz/(MV/m)$^2$.

In addition to these measurements, some work has been done to develop active compensation for these cavities. Figure 5 shows the detuning with active compensation for the SSR1 type cavities tested in a single cavity test stand. Here we see that while the RMS detuning is within the specification, the peak detuning is still outside the specification. While these results are optimistic there is still a bit of work to be done in order to ensure that the resonance control system will be able to meet the specifications.

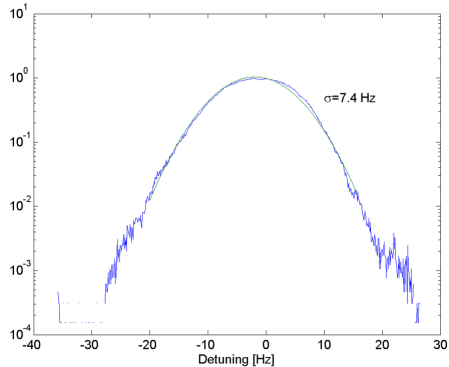

Figure 5: Active compensation test results from 325 MHz test stand

## LLRF SYSTEM FOR PIP-II

The LLRF system for PIP-II utilizes modular hardware from the rack level down to the chassis level. The racks are organized into four-cavity stations. Each station has two 8-channel down-converters, one 4-channel up converter, two 2-cavity controller modules, one 4-cavity resonance controller module and a LO/Clock distribution chassis. Figure 6 shows a schematic layout of the 4-cavity RF station.

The LLRF system is organized in a group of up to four cavities serviced by one rack of electronics as shown in Figure 6 with each module described in detail below. The group of four cavities allows for an economy of scale in the hardware design while keeping cable runs as short as possible. The signal path is kept as direct as possible with cables from the accelerator tunnel brought directly to the 8-channel down-converter.

The down-converter is used to translate the RF signals at the various frequencies across the accelerator to a common intermediate frequency (IF) of 20 MHz. In order to achieve our regulation requirements the down-converter must be, low-noise (-158dBc), have high channel isolation (greater than 80dB), phase stability to within 0.005 degrees above 1 Hz, and amplitude stability 0.005% above 1 Hz.

The IF is digitized by a high-speed Analog to Digital Converter (ADC). The ADC must have a sample rate greater than 94 MS/s, and goals of achieving -155dBc noise power density and less than 200 ns of signal latency to maximize control loop bandwidth. There are several ADCs on the market that achieve these specifications; the present choice is the AD9653 from Analog Devices. The digitizer board has 8 channels that are sent to the Field Programmable Gate Arrays (FPGA).

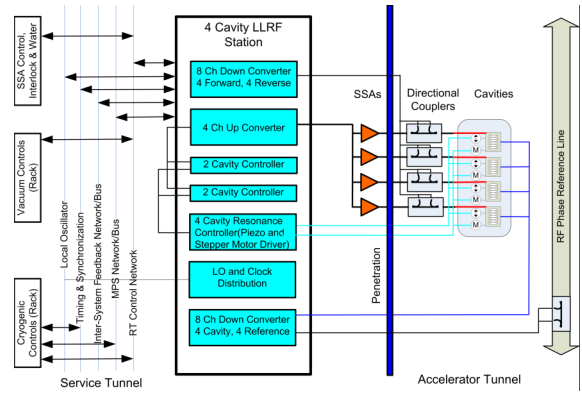

Figure 6: Schematic of the 4-cavity LLRF station for PIP-II

The FPGA processes the digitized signals for both read-back and for input to the controller. The controller will have a closed loop bandwidth of ~30 kHz and is designed to support both CW and pulsed operation. Each controller will have the ability to run in a self-excited loop where the drive frequency tracks the cavity, and in a generator driven mode where the output is at a fixed frequency. It is expected that each cavity will be operated CW during some part of commissioning and tests. CW operation typically requires startup in a self-excited loop with a transition to a generator driven loop to align with the beam phase. For cavities with a low bandwidth, pure pulsed operation requires a complex phase trajectory program to fill the cavity with energy and stay in regulation for the pulse. A prototype system is under development for the PIP-II injector test (see Figure 7). It includes the LLRF four Station Field Controller Module shown in Figure 3.85. The LLRF system provides both RF waveforms and sampled values to the control system that are calibrated and highly linear. They best represent the cavity field and directional RF signals and will be used for all data analysis.

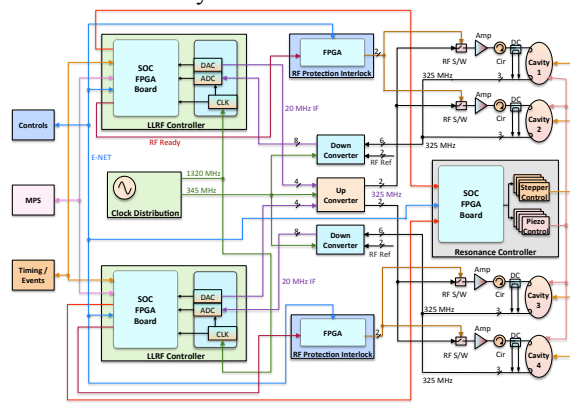

Figure 7: Control signal block diagram

## CHOPPER PROGRAM GENERATOR

The LLRF system will also generate the waveforms needed for the beam chopper[*] and the 44.705 MHz RF signal for the Booster to lock to during the ~1 millisecond injection period. The waveforms require complex pre-distortion for the chopper amplifiers, which is better implemented with the entire waveforms calculated and played out from pre-calculated tables. There are several advantages to waveform tables: repeatability from pulse to pulse, local storage of beam waveforms in LLRF and instrumentation systems, and the ability to make small corrections utilizing adaptive feed-forward. A multi-channel 4 GSPS arbitrary waveform generator is specified for this purpose. Figure 8 shows a schematic diagram of the chopper waveform signal chain.

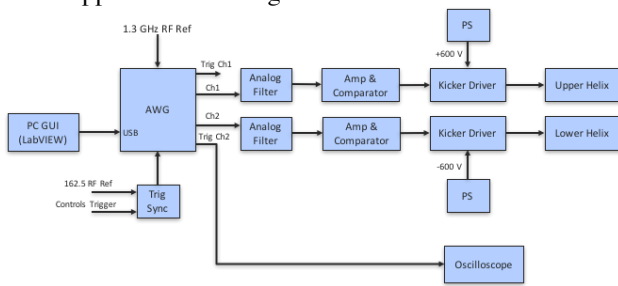

Figure 8: Block diagram of the chopper waveform generator

At present the chopper waveform generator has been tested in the injector test stand for PIP-II and has demonstrated the ability to do bunch-by-bunch chopping of the beam out of the RFQ.

## HARDWARE STATUS

Currently we have built and tested prototype up-converter and down-converter modules for PIP-II. The four channel up converter takes in a 20 MHz IF signal at -2dBm max and can generate either 162.5 MHz, 325 MHz, or 650 MHz depending on the LO. The output of the up converters are +11 dBm max. These modules are highly linear for all frequencies. Figure 9 shows the linearity of the up converter for each frequency. The channel-to-channel isolation for these modules is better than 88 dB and the spurious signal suppression is better than 80 dB.

---

[*] The beam chopper removes bunches on the boundary of RF buckets and forms a 3 bunch long extraction gap.

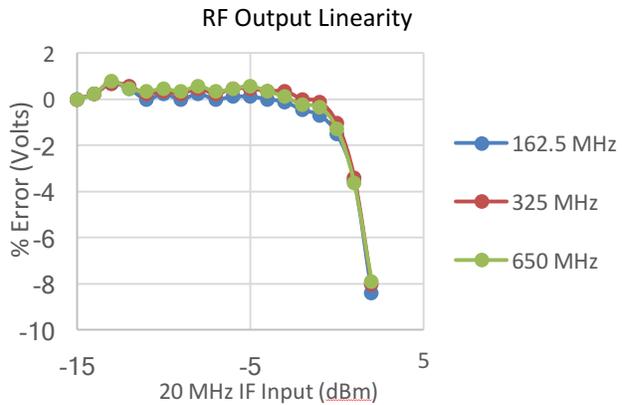

Figure 9: Linearity measurement of the 4-channel up-converter

We have also tested the linearity and isolation of the prototype 8-channel down-converter. These modules cover the 162 MHz to 650 MHz RF input range, with LO inputs 20 MHz above the RF frequencies, downcoverting to a 20 MHz IF for all cavity types. They have a linearity better than 1% for an RF input up to 10 dBm. Figue 10 shows the linearity of these modules.

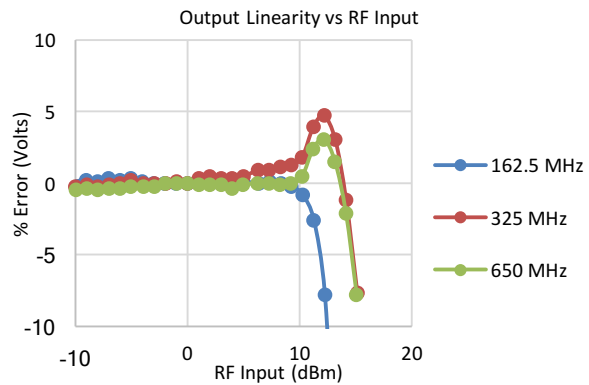

Figure 10: Linearity measurement of the 8-channel downconverter

The channel-to-channel isolation for these modules is better than 82 dB and the noise output floor is -161 dBc/Hz. The RF hardware development for PIP-II is making good progress and we have shown very high quality analog signal processing results.

## PROGRESS OF THE IIFC COLLABORATION

The PIP-II project also includes an international collaboration with the Department of Atomic Energy in India. The goal of this collaboration is a joint design of LLRF hardware that will be constructed by the DAE and tested at Fermialb. Thus far the collaboration has been successful with seven joint specifications approved by both parties and several more on the way. In addition we have prototype hardware under construction by the DAE that will hopefully be tested at FNAL in the near future.

Figures 11 and 12 show photos of the RF hardware currently under construction in India.

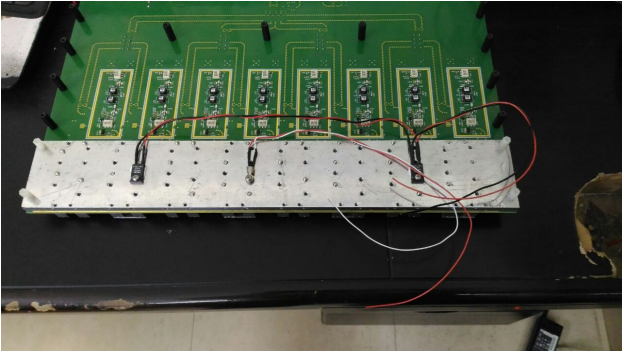

Figure 11: Eight channel receiver constructed by the DAE

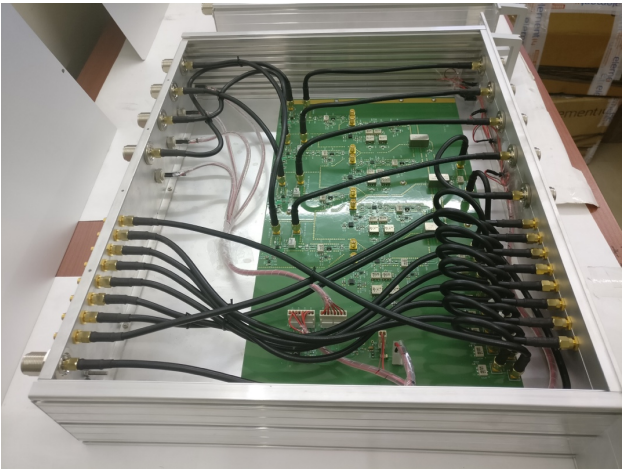

Figure 12: Four channel up-converter constructed by the DAE

## CONCLUSIONS

The LLRF system for the PIP-II accelerator has many challenges, however we are optimistic that with the appropriate development time we will be able to overcome these challenges. We have confidence that we will be able to meet the field control requirements and, consequently, the energy and phase stability requirements. We are cautiously optimistic that the resonance control efforts will be able to deliver a system that meets the required detuning specifications. Additionally we are making good headway on architecture and hardware designs that will pave the way for smooth operations. Our modular approach to hardware, software, and firmware should allow for easy transition to the next generation of systems as technology matures.